\documentclass[%
 reprint,
 superscriptaddress,
 amsmath,amssymb,
 prx
]{revtex4-2}

\usepackage{gensymb}
\usepackage{graphicx}
\usepackage{dcolumn}
\usepackage{bm}
\usepackage{braket}
\usepackage{comment}

\begin{document}

\preprint{APS/123-QED}

\title{A hybrid quantum photonic interface for $^{171}$Yb solid-state qubits}

\affiliation{Thomas J. Watson, Sr., Laboratory of Applied Physics, California Institute of Technology, Pasadena, CA, USA}
\affiliation{Institute for Quantum Information and Matter, California Institute of Technology, Pasadena, CA, USA}
\affiliation{Kavli Nanoscience Institute, California Institute of Technology, Pasadena, CA, USA}
\affiliation{Division of Physics, Mathematics and Astronomy, California Institute of Technology, Pasadena, CA, USA}
\author{Chun-Ju Wu$^{1,2,3,4,}$}
\altaffiliation{These authors contributed equally to this work.}
\author{Daniel Riedel$^{1,*,}$}
 \altaffiliation{Present address: AWS Center for Quantum Networking, Boston, Massachusetts, USA}
\author{Andrei Ruskuc$^{1,2,3}$}
\author{Ding Zhong$^{1,2,3}$}
\author{Hyounghan Kwon$^{1,3,}$}
\thanks{Present address: Center for Quantum Information, Korea Institute of Science and Technology, Seoul, Republic of Korea}
\author{Andrei Faraon$^{1,2,3,}$}
\email{Faraon@caltech.edu}

\date{\today}

\begin{abstract}

$^{171}$Yb$^{3+}$ in YVO$_4$ is a promising candidate for building quantum networks with good optical addressability, excellent spin properties and a secondary nuclear-spin quantum register. However, the associated long optical lifetime necessitates coupling to optical resonators for faster emission of single photons and to facilitate control of single $^{171}$Yb ions. Previously, single $^{171}$Yb ions were addressed by coupling them to monolithic photonic crystal cavities fabricated via lengthy focused ion beam milling. Here, we design and fabricate a hybrid platform based on ions coupled to the evanescently decaying field of a GaAs photonic crystal cavity. We experimentally detect and demonstrate coherent optical control of single $^{171}$Yb ions. For the most strongly coupled ions, we find a 64 fold reduction in lifetime. The results show a promising route towards a quantum network with $^{171}$Yb:YVO$_4$ using a highly scalable platform.

\end{abstract}

\maketitle

\section{\label{sec:introduction} Introduction}
Transmitting quantum information and distributing entanglement through quantum networks are essential components in quantum technology and have applications in quantum communication and distributed quantum computing \cite{kim08,weh18}. Building a scalable optical quantum network requires nodes with lifetime limited optical transitions, efficient optical interfaces, long spin coherence times, high-fidelity spin and optical control, and multi-qubit accessibility at each node. Among different platforms, optically addressable solid-state spin qubits are promising candidates due to the possibility for integration with nanofabricated devices leading to scalability \cite{aws18,wol21}. Possible candidates include nitrogen-vacancies \cite{pom21,her22}, silicon-vacancies \cite{zha20,sta22} and tin-vacancies  \cite{rug21,deb21} in diamond, color centers in SiC \cite{wol20,luk20}, defects in silicon \cite{hig22}, and rare-earth ions in crystals \cite{kin20,che20}. Rare-earth ions have been shown to possess long spin and optical coherence times in various hosts and will be the focus of this work \cite{ste22,zho15}.

\begin{figure*}
\includegraphics{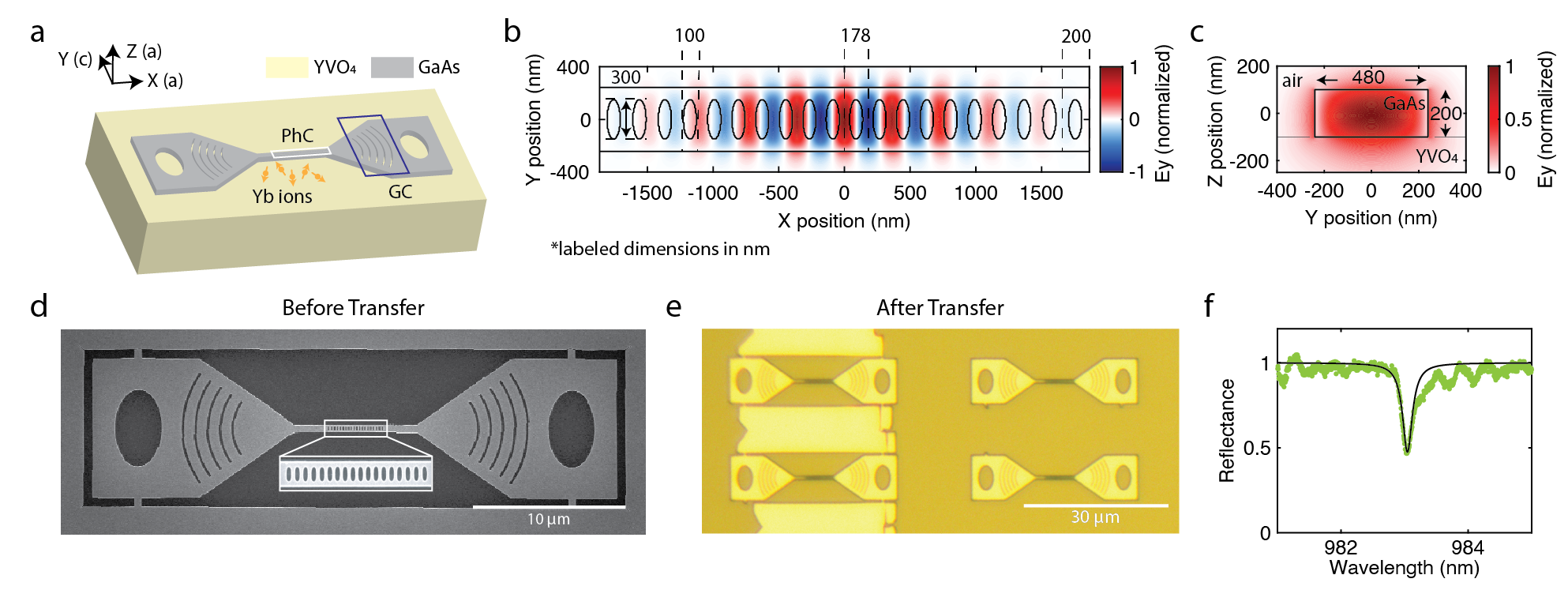}
\caption{\label{fig:1}Hybrid device platform for coupling to $^{171}$Yb:YVO$_4$. (a) The device schematic. The GaAs photonic crystal cavity and YVO$_4$ crystal are shown in gray and yellow, respectively. Yb ions are shown with orange arrows and are located inside YVO$_4$. The crystal axes are shown next to the coordinate axes. The grating coupler (blue rectangle) is used for coupling light to a free-space setup. The photonic crystal cavity is indicated with a white rectangle. (b), (c) Simulated E$_y$ field of the fundamental TE mode in the XY and YZ planes, respectively. Note the evanescent decay of electric field inside YVO$_4$. (d) Scanning electron microscope image of the suspended GaAs photonic crystal cavity before transferring onto YVO$_4$. (e) Optical image of the transferred device. The darker yellow substrate under the GaAs photonic crystal cavities is YVO$_4$. (f) Cavity reflection spectrum after transferring onto YVO$_4$. Data are shown in green dots and overlayed with a fitted black line, which gives a quality factor of 5300.}
\end{figure*}

In rare-earth ion platforms, coherent 4f-4f optical transitions are only weakly allowed inside crystals, therefore it is essential to couple ions to cavities with large quality factors and small mode volumes to increase the emission rate through Purcell enhancement. However, the nanofabrication of monolithic nanophotonic cavities for common rare-earth host materials is limited due to the unavailability of high-quality thin films and selective etching chemistries (aside from a few examples \cite{dut20,xia22,gri22}). Here we demonstrate a hybrid platform comprising a separately fabricated photonic crystal cavity that is subsequently transferred onto the host crystal (Figure \ref{fig:1}a) \cite{dib18,hua21}. This platform possesses a smaller mode volume compared to Fabry-Perot microcavities \cite{rie17,mer20}, and unlike the previous approach of focused ion beam milling \cite{kin20,zho16}, doesn't lead to crystal damage. 

Specifically, we couple $^{171}$Yb$^{3+}$ ions inside YVO$_4$ to an external GaAs photonic crystal cavity. Previously, we demonstrated that these ions have spin coherence times exceeding 10 ms, over 99.9$\%$ single-qubit gate fidelities, 95$\%$ optical readout and initialization fidelity, and additional nuclear spin qubit control which are essential properties for building a quantum network \cite{kin20,rus22}.
GaAs is transparent at 984.5 nm: the optical transition wavelength of $^{171}$Yb:YVO$_4$, and high quality factor photonic crystal cavities have been fabricated at this wavelength using ebeam lithography, thus enabling mass production \cite{mid15,kur20}. In the following sections of the paper, we will present the design and fabrication of these hybrid devices, and experimentally demonstrate optical control of single $^{171}$Yb ions.

\section{Device Design and Fabrication}
 Yb ions inside an a-cut YVO$_4$ crystal are coupled to a 1-dimensional photonic crystal cavity with fundamental TE mode \cite{qua11}, where the electric field is mostly aligned with the c-axis of the crystal. The photonic crystal design is based on unit cells with elliptical holes to engineer a bandgap at 984.5 nm. A localized cavity mode is formed by quadratically tapering 20 central periods of a 44-period photonic crystal (Figure \ref{fig:1}b). This is optimized to reduce radiation loss while maintaining a small mode volume. We engineer preferential cavity--waveguide coupling on one side by removing 11 mirror periods. The thickness of GaAs (200 nm) is chosen to enable high field penetration into YVO$_4$ without decreasing the quality factor significantly. Figures \ref{fig:1}b,c show the simulated y-directed electric field ($E_y$) that will be aligned with the Yb optical dipole moment along the c axis, along with the pertinent cavity dimensions. The simulated squared electric field magnitude ($\left|E^2 \right|$) at the YVO$_4$--GaAs interface is 40$\%$ of the maximum $\left|E^2 \right|$ inside the GaAs layer. The effective mode volume is $\sim$1.7 $(\lambda/n_{\rm{YVO_4}})^3$ (normalized according to the strongest electric field in YVO$_4$). $\left|E^2 \right|$ evanescently decays with distance from the interface, reducing by half every 30 nm in YVO$_4$. To couple light from the cavity into a fiber, we utilize a fully-etched grating coupler \cite{miy17}. The grating coupler efficiency is optimized by modifying the grating dimensions leading to a simulated efficiency of $\sim$25$\%$. 
 
Devices are fabricated on an epitaxial GaAs wafer using a standard electron-beam lithography
procedure with an acceleration voltage of 100kV (Raith EPBG 5000) \cite{mid15}. A 500-nm positive-tone resist (ZEP) is patterned and subsequently developed using n-Amyl acetate
(ZED, Zeon corp.). The resist serves as the etch mask for an optimized inductively-coupled
plasma reactive-ion etcher with Cl$_2$-Ar chemistry. Devices are undercut by etching an AlGaAs sacrificial layer in HF and cleaned subsequently using H$_2$O$_2$ and KOH to remove any residues (Figure \ref{fig:1}d).
Finally, suspended GaAs photonic crystal cavities are transferred onto the YVO$_4$ surface using a stamping technique \cite{dib18}. Devices are aligned with the a-axis and are perpendicular to the c-axis. We use a polydimethylsiloxane stamp covered by a thin film of polycarbonate to pick up devices and subsequently release them onto YVO$_4$ using a transfer stage. The success rate of this procedure is  over 90$\%$. The YVO$_4$ crystal used in this experiment was cut perpendicular to its a-axis and polished from an undoped boule (Gamdan Optics) with an Yb concentration of 0.14 ppm. The optical image after the transfer is shown in Figure \ref{fig:1}e; the device used in these measurements has a quality factor of 5300 after transferring onto YVO$_4$ (Figure \ref{fig:1}f).

\begin{figure}
\includegraphics{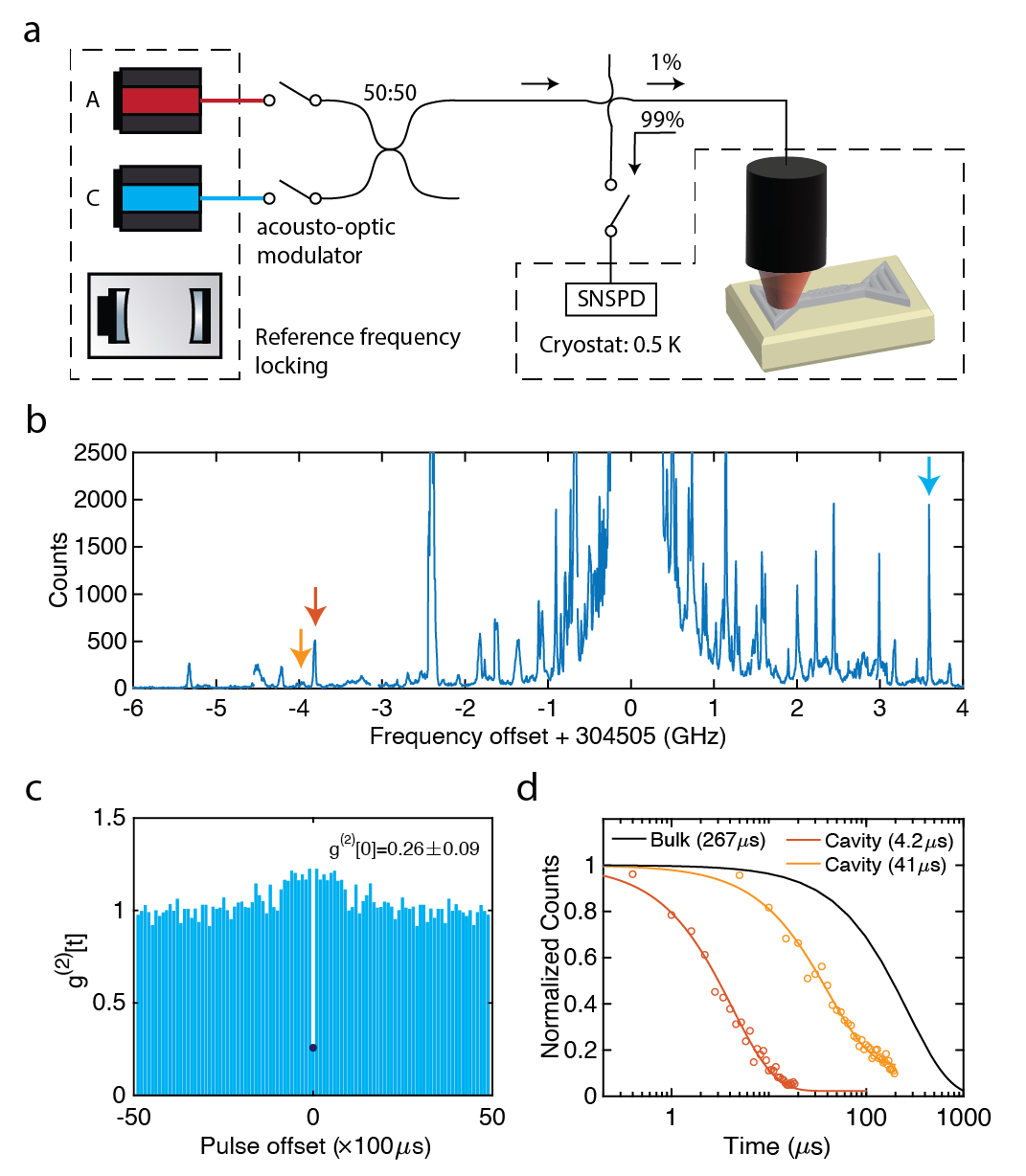}
\caption{\label{fig:2}Experimental setup and detection of single Yb ions. (a) Experimental setup. Optical pulses from two frequency locked lasers, labeled A and C according to the driven optical transition (Figure \ref{fig:3}a), are sent to the device located in a $^{3}$He fridge at 0.5 K. Emitted photons are routed to an SNSPD in the same cryostat. (b) Resonant photoluminescence spectrum. The large emission peak at 304505 GHz corresponds to Yb isotopes with no nuclear spin. Isolated peaks inside the spectrum are mostly single Yb ions, the arrows indicate ions measured in subsequent experiments. Data was taken with 10 second integration time and 50 kHz repetition rate for each point. (c) Pulsed g$^{(2)}$(t) measurement of the ion indicated with a blue arrow shows g$^{(2)}$(0)=$0.26\pm0.09$. (d) Lifetime measured through time-resolved photoluminescence of ions indicated by red ($4.2\pm0.1$ $\mu s$) and orange ($41\pm2$ $\mu s$) arrows are shown with corresponding colors. Experimental data are shown with dots and overlayed with fitted exponential decays. Bulk lifetime is shown with a black line.}
\end{figure}

\section{Detection of Single Yb Ions in YVO$_4$}
Experiments are performed in a Bluefors $^3$He fridge (LD-He250) at 0.5 K and zero magnetic field. The experimental setup is depicted in Figure \ref{fig:2}a, where acousto-optic modulated laser pulses are sent through a 99:1 beamsplitter and focused onto the grating coupler through an aspheric lens doublet. The ion emission is sent back through the same fiber, and 99$\%$ of the light is directed to a superconducting nanowire single photon detector (SNSPD). The light coupling is optimized using an x-y-z nanopositioner (Attocube), and the device resonance is tuned to the ion emission frequency through nitrogen condensation.

Figure \ref{fig:2}b shows a resonant pulsed photoluminescence spectrum, where ion emission has been distinguished from the excitation in the time domain. The zero frequency offset corresponds to emission from Yb isotopes with no nuclear spin. At zero magnetic field, these isotopes contain degenerate Kramers doublets in both the ground and excited states \cite{kin18}. The arrows indicate the ions used in this work.

To demonstrate single ion addressability, a single detector pulsed g$^{2}$(t) measurement was performed on the ion indicated by the blue arrow (Figure \ref{fig:2}c). The resulting g$^{2}$(0) is $0.26\pm0.09$, lower than the two-ion limit of 0.5. The g$^{2}$(t) measurement demonstrates a weak bunching feature, likely caused by spectral diffusion of the optical transition. This enables us to postulate a spectral diffusion correlation time-scale of about 1 ms \cite{sal10}. 
The ion indicated by the red arrow has the shortest lifetime in this sample, measured to be $4.2\pm0.1$ $\mu s$ (Figure \ref{fig:2}d). %This corresponds to a Purcell enhancement factor of 183, and 
This corresponds to a lifetime reduction of 64 times. 

\begin{figure*}
\includegraphics{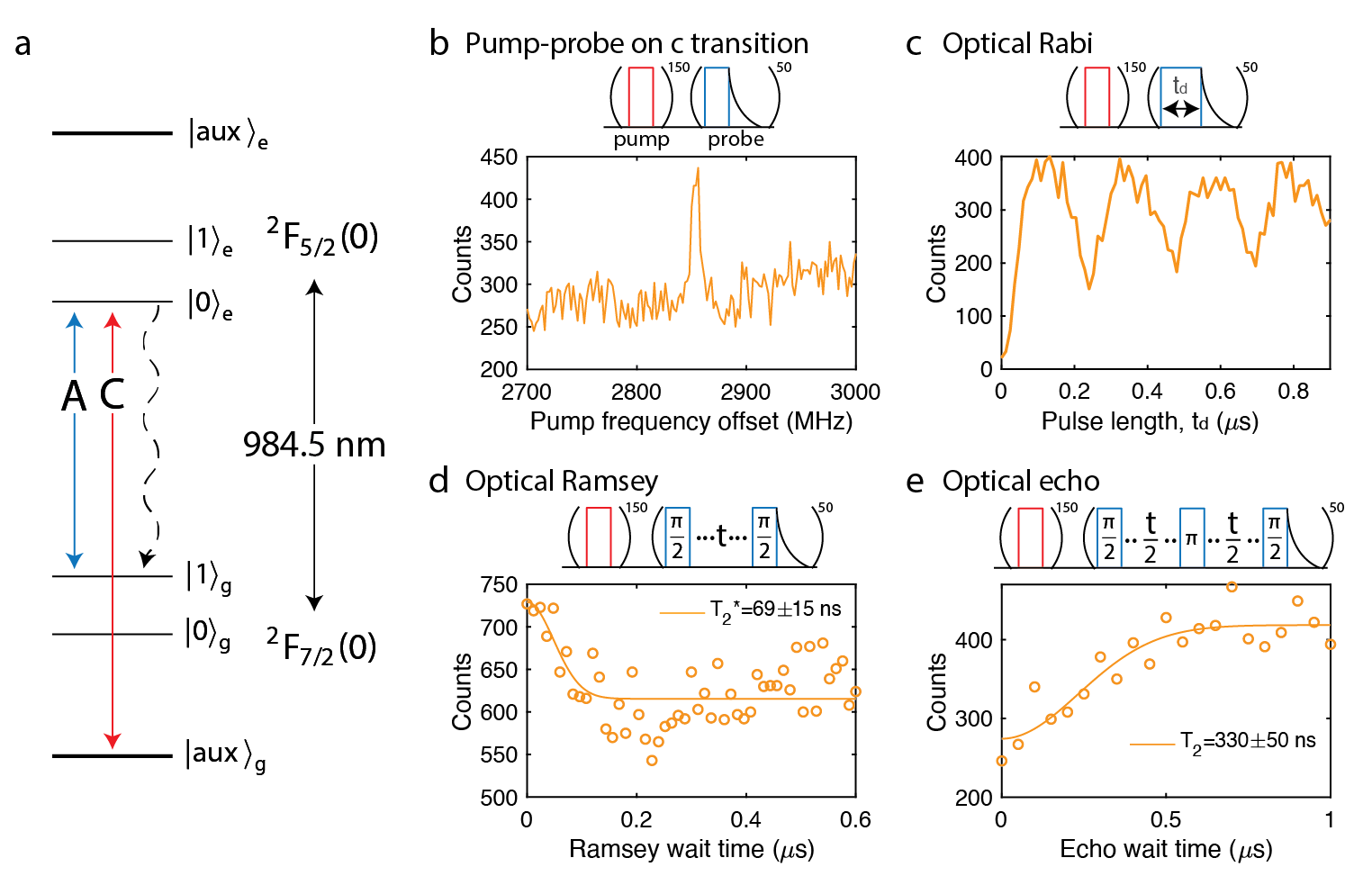}
\caption{\label{fig:3}Coherent optical control of a single $^{171}$Yb ion. (a) Energy level structure of $^{171}$Yb inside YVO$_4$ at zero magnetic field. Transition A (cavity enhanced) is shown in blue; transition C (non-enhanced) is shown in red. Cavity enhanced decay is shown with a dashed arrow. $\ket{0}_g$, $\ket{1}_g$, $\ket{0}_e$, $\ket{1}_e$ are magnetic field insensitive forming clock transitions. (b) Pump-probe measurement of the $^{171}$Yb optical transitions. The pump frequency is varied around the C-transition with probe frequency fixed to the A-transition. In subsequent measurements, the ion is partially initialized by repeatedly pumping on this transition. (c) Optical Rabi oscillation on the A-transition. (d) Optical Ramsey measurement on the A-transition gives $T_2^*=69\pm15$ ns. (e) Optical echo measurement on the A-transition gives $T_2=330\pm50$ ns.}
\end{figure*}

\section{Coherent optical control of a single $^{171}$Yb ion}

While Yb isotopes without nuclear spin show bright emission without initialization, they suffer from poor coherence times  due to first order susceptibility of transition frequencies to magnetic fields. Hereafter, we focus on $^{171}$Yb ions which have an additional 1/2 nuclear spin giving a zero field hyperfine energy level structure shown in Figure \ref{fig:3}a. These ions exhibit optical and spin clock transitions between the $\ket{0}_g$, $\ket{1}_g$, $\ket{0}_e$, and $\ket{1}_e$ levels, which are first order insensitive to magnetic field noise yielding enhanced coherence properties \cite{kin20}. The A-transition has its dipole moment along the c-axis of the crystal;  it is copolarized with, and enhanced by, the cavity. Furthermore, it has no overlap with other Yb isotopes, making it suitable for optical readout. The C-transition has dipole moment along the a-axis of the crystal and is used for optical pumping and initialization.

To verify that the ion indicated by the orange arrow is $^{171}$Yb (Figure \ref{fig:2}b), we performed a pump-probe measurement, where we varied the pump laser frequency and measured the photoluminescence on the A-transition. When the pump laser is resonant with the C-transition, the ion will be partially initialized into $\ket{1}_g$ and have brighter emission (Figure \ref{fig:3}b). This initialization is performed prior to all subsequent measurements. 

Resonant photoluminescence with varied excitation pulse length on the A-transition shows optical Rabi oscillation (Figure \ref{fig:3}c). In this measurement, initialization is periodically applied after each sequence of 50 readout pulses. Combined with the limited cyclicity of the readout transition, this acts to saturate the photon counts and create a flat top feature. This ion has a lifetime of $41\pm2$ $\mu$s, leading to an A-transition cyclicity of 10 \cite{kin20}. Characterization of the optical coherence properties is first performed with an optical Ramsey measurement, which includes two optical $\pi/2$ pulses with varied separation. We measured $T_2^*=69\pm15 $ ns corresponding to an effective linewidth of 4.6 MHz. The short-timescale optical frequency stability is measured using an echo consisting of an additional intermediate optical $\pi$ pulse to rephase the coherence, yielding $T_2=330\pm50$ ns.

In these experiments, we were unable to perform optical coherence measurements for some of the ions that have shorter lifetime. Ions in these hybrid devices with larger Purcell factors are closer to the surface and therefore more susceptible to surface defects. The resulting charge noise leads to excess spectral diffusion and deteriorated coherence properties. By contrast, in previously studied monolithic devices, well coupled ions showed lifetime limited coherence since they were centrally located and farther from surfaces. The next step would be to characterize ion statistics with different Purcell factors and understand the limiting factor for the optical coherence in these hybrid devices. 

\section{Conclusion and Outlook}
In this work, we fabricate suspended GaAs photonic crystal cavities and transfer them onto YVO$_4$ with a $\sim$ 90$\%$ success rate. We use a cavity with a quality factor of 5300 to address single $^{171}$Yb ions, show a lifetime reduction of 64 times, and measure their optical coherence properties. 

In the future, developing coherent microwave control of the spin transition will be essential. To improve the platform and increase count rates, refining fabrication of the photonic crystal cavity through surface passivation and further optimization of the lithographic procedure are crucial \cite{guh17,kur20}. Switching to a shallow-etched grating coupler can also increase the waveguide to free-space coupling efficiency \cite{zhou18}. If the optical coherence is limited by surface proximity, we could dope Yb ions at different depths and characterize their properties \cite{ste22}. Given the fabrication capability and the flexibility of the hybrid GaAs platform combined with the excellent properties of $^{171}$Yb:YVO$_4$, we think this approach is a promising route to build a quantum network.

\begin{acknowledgments}
This work was funded by Office of Naval Research award no. N00014-19-1-2182, National Science Foundation award no. 1936350, DOE-QIS program (DE-SC0019166), and Northrop Grumman. C.-J.W. acknowledges the support from the Taiwanese Government scholarship. D.R.
acknowledges support from the Swiss National Science
Foundation (Project No. P2BSP2\_181748). A.R. acknowledges the support from Eddleman Graduate Fellowship. D.R. contributed to this work prior to joining AWS. The device nanofabrication was performed in the Kavli Nanoscience Institute at the California Institute of Technology. We thank Joonhee Choi, Jake Rochman, Ioana Craiciu, Tian Xie, Mi Lei, Rikuto Fukumori, and Helena Guan for discussion, and Matt Shaw for help with superconducting photon detectors. 
\end{acknowledgments}

\appendix

\nocite{*}

\bibliography{reference}
\end{document}